\documentclass{article}

\usepackage[final]{neurips_2026}
\usepackage[utf8]{inputenc}
\usepackage[T1]{fontenc}
\usepackage{amsmath,amssymb,amsfonts}
\usepackage{booktabs}
\usepackage{microtype}
\usepackage{xcolor}
\usepackage{enumitem}
\usepackage{hyperref}
\usepackage{url}
\usepackage{graphicx}
\usepackage{xspace}

\newcommand{\stabilityarc}{\textsc{StabilityArc}\xspace}

\title{StabilityArc: Decoding Protein Sequence Embeddings into Generalizable Stability Landscapes}

\author{
    Aaron L. Feller\\
    The University of Texas at Austin\\
    Interdisciplinary Life Sciences\\
    \texttt{aaron.feller@utexas.edu}
    \And
    Andrew D. Ellington\\
    The University of Texas at Austin\\
    Interdisciplinary Life Sciences\\
    \texttt{andy.ellington@austin.utexas.edu}
    \And
    Claus O. Wilke\\
    The University of Texas at Austin\\
    Integrative Biology\\
    \texttt{wilke@austin.utexas.edu}
    }

\makeatletter
\def\@trackname{}
\makeatother

\begin{document}

\maketitle

\begin{abstract}
Every protein has a unique stability landscape, but the physical consequences of mutation are governed by recurring biochemical constraints. We test whether a shared decoder, trained on measurements from diverse proteins, can interpret these constraints in an unseen target, enabling cross-protein transfer for initial experimental round prescreening. We present \stabilityarc, which maps frozen ESMC-600M residue representations through a shared RoPE transformer to an $L\times20$ matrix of substitution effects; a symmetric, contact-aware residual aids in predicting epistasis in simultaneous substitutions. In 66 strict leave-one-protein-out evaluations covering 134,794 ProteinGym variants, \stabilityarc achieves 0.7134 Spearman correlation, exceeding the strongest zero-shot baseline, ProSST-2048 (0.6526), by 0.0608. We further explore the utility of this method by providing the score as a prior for Kermut, achieving Spearman correlation of 0.8280 across three supervised split schemes, improving on Kermut's reported 0.8167.
\end{abstract}

\section*{Introduction}

The majority of proteins exist in a carefully folded organization that is both unique and necessary to confer the specific function of that protein \citep{martin1998protein}. When engineering a protein, our goal is to modify the sequence of that protein to enhance protein function, while avoiding disruption of the often fragile form that is required for activity \citep{listov2024opportunities}. Often, we do not have available data, and the evaluation of that protein is very low throughput, requiring high cost and time. However, there are many proteins that have been evaluated for stability in high throughput manner. Recent large-scale measurements from deep mutational scanning (DMS) have been used to map mutations to their effect on protein's stability \citep{fowler2014dms,riesselman2018dms,tsuboyama2023stability}. 

A stability landscape is highly local to a sequence and its folded state: the same substitution can be benign in one context and damaging in another \citep{karagol2025co}. Yet the physical mechanisms that penalize mutations, including packing frustration, unsatisfied charge, and solvent exposure, recur universally \citep{otwinowski2018biophysical}. If a pre-trained embedding provides the structural and biochemical prior for the wild-type environment, a shared decoder, trained across many proteins, may be able to map amino acid representations to the impact of substitutions on stability shifts. The evaluation should be simple: can a single decoder trained across multiple protein stability landscapes transfer to an out-of-distribution protein?

\begin{figure}[t]
  \centering
  \IfFileExists{figures/fig1.png}{%
    \includegraphics[width=\linewidth,keepaspectratio]{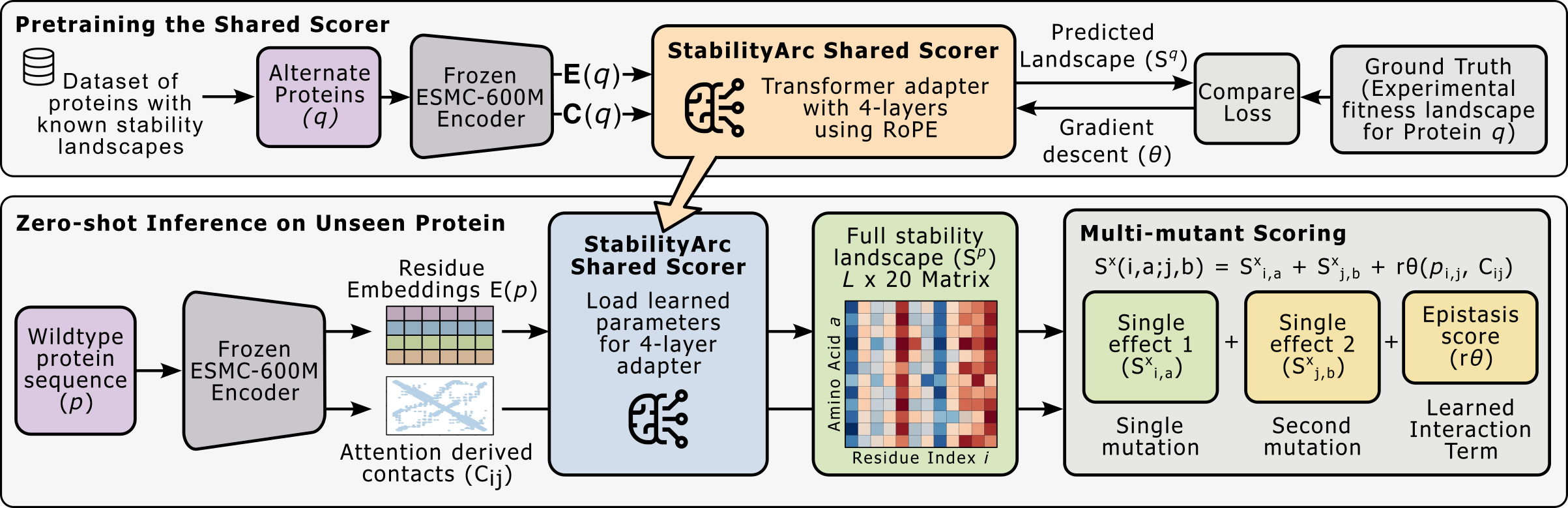}%
  }{%
    \fbox{\parbox[c][1.45in][c]{0.94\linewidth}{\centering
      \textbf{Figure 1 placeholder.}\\[0.35em]
      Wild type $\rightarrow$ frozen ESMC $\rightarrow$ shared RoPE scorer $\rightarrow$ $L\times20$ landscape;\\
      contact-aware composition; held-out score $\rightarrow$ Kermut prior.
    }}%
  }
  \caption{Cross-protein learning and deployment of \stabilityarc. During training (top), alternate proteins ($q$) are passed through a frozen ESMC-600M encoder to train the shared 4-layer RoPE transformer adapter against ground-truth experimental fitness landscapes. All labels from the target protein are strictly withheld. At deployment (bottom), the target wild-type sequence ($p$) is encoded to provide structural representations and contact maps, which the learned adapter decodes into a complete $L\times20$ single-substitution stability landscape. For simultaneous substitutions, a multi-mutant scoring module (right) combines these single effects with a learned epistasis interaction term using attention-derived contacts.}
  \label{fig:overview}
\end{figure}

\stabilityarc (Figure~\ref{fig:overview}) explores this potential; frozen ESMC-600M embeddings and attention-derived contacts provide a label-free description of the target wild type \citep{candido2026worldmodel,rao2021transformerplm}. A shared four-layer transformer decodes that description into effects for all 20 substitutions at every site, and a symmetric contact-aware residual accounts for epistasis in double mutants. During held-out evaluation, every DMS label from the target protein is excluded; the target supplies sequence-derived features only.

We evaluate the resulting landscape at two points in an engineering cycle. Before target measurements, we use \stabilityarc to rank variants under strict protein holdout. After measurements arrive, the score from \stabilityarc is used as a prior in Kermut \citep{groth2024kermut}, a Gaussian-process (GP) model that revises the landscape using target-assay data, structure, and variant embeddings. Ultimately, we demonstrate that a robust cross-protein prior can both succeed at predicting mutation effects in a blind manner and can seamlessly integrate with target-specific optimization once experimental data becomes available.

\section*{Method}

\paragraph{Shared wild-type decoder.}
Let $x=(x_1,\ldots,x_L)$ be a wild type sequence and let $E(x)\in\mathbb{R}^{L\times d}$ be its frozen ESMC-600M residue embeddings. We train a learned projection using four pre-norm transformer layers \citep{vaswani2017attention} with rotary position embeddings \citep{su2024roformer} and eight attention heads to produce residue states $H\in\mathbb{R}^{L\times256}$. A linear head applied to the embedding at a given position gives:
\begin{equation}
S^x=f_\theta(E(x))\in\mathbb{R}^{L\times20},
\label{eq:landscape}
\end{equation}
where $S^x_{i,a}$ is the predicted effect of substituting amino acid $a$ at site $i$. Parameters $\theta$ are shared across proteins.

\paragraph{Multiple mutations with contact-aware composition.}
To predict the effect of a double mutation, the model sums the individual substitution scores ($S_1$ and $S_2$) and adds a learned interaction term ($r$):
\begin{equation}
\hat y = S_1 + S_2 + r.
\label{eq:pair}
\end{equation}
The term $r$ (further explained in Appendix A.3) calculates how the two mutations influence each other based on their sequence distance and their likelihood of physical contact in the folded structure. Following prior work showing that protein language models capture structural information \citep{rao2021transformerplm}, we extract these contact probabilities directly from the frozen ESMC attention maps rather than relying on experimental coordinates.

\paragraph{Learning objective.}
Training combines Huber loss with a pairwise ranking loss. For each ranking-loss evaluation, we select a batch of 1,024 random pairs of variants and train their predicted ordering to match their measured ordering. Single-mutant labels are used directly. Double-mutant labels are centered and scaled within each assay using the median and median absolute deviation of that assay's single-mutant labels.

\paragraph{Supervised prior.}
The Gaussian process in Kermut builds its baseline prediction by taking a predicted score as a prior and adjusting it with a simple linear equation \citep{groth2024kermut}. It pairs this with a covariance function that combines structural mutation similarity and an RBF kernel over ESM-2 variant embeddings. For a target protein $p \notin P$, we replace only its mean input:
\begin{equation}
m(v)=b+\beta s_{\stabilityarc}^{p}(v),
\label{eq:prior}
\end{equation}
where $s_{\stabilityarc}^{p}$ is the predicted stability for the target protein, generated by the model trained on the remaining dataset. The GP still learns from the four allowed training folds of the target assay, separating cross-protein transfer in the prior from within-assay posterior adaptation.

\section*{Experimental Setup}

\paragraph{Data and hold-one-protein-out protocol.}
We use 66 ProteinGym stability assays containing 134,794 variants: 68,855 singles and 65,939 doubles. We train 66 outer models, each excluding all labels for one protein. The held-out wild-type sequence and frozen contacts are available at inference. Singles and doubles are concatenated within each assay, one Spearman correlation is computed, and the unweighted mean over assays is reported. Every model is trained for 40 epochs with AdamW, learning rate $10^{-4}$, weight decay $10^{-3}$, and up to 40,000 measured singles and doubles per training assay.

\paragraph{Comparisons.}
For cross-protein comparison, we include methods with complete score coverage for the same variants in all 66 assays and report pooled, single-only, and double-only macro Spearman. The inputs for various models differ: e.g. ProSST uses AlphaFold2-derived structure tokens, whereas \stabilityarc uses ESMC-600M embeddings from the wild-type sequence \citep{candido2026worldmodel}. For supervised adaptation, we follow ProteinGym's \citep{notin2024proteingym} official five-fold random, modulo, and contiguous splits, yielding $66\times3\times5=990$ GP fits. We calculate each assay's Spearman $\rho$ across all available measurements.

\section*{Results}

\paragraph{Cross-protein transfer.}
For prediction of protein stability, each assay is scored by a model trained without that protein, so the result measures cross-protein transfer rather than interpolation from its labels. In this schema, \stabilityarc reaches 0.7134 all-variant macro Spearman (Table~\ref{tab:zero-shot}), 0.0608 above the leading model, ProSST-2048. \stabilityarc also leads the single- and double-mutant ranking individually, where again, ProSST-2048 previously held the strongest overall baseline. 

In a second cross-protein evaluation on the dataset T2837 \citep{ouyang2023predicting}, \stabilityarc improves recall and accuracy (Supplementary Table~\ref{tab:t2837-supp}) but shows decreased Spearman, AUROC, and precision when compared to StabilityOracle \citep{diaz2024stability}.

\begin{table}[t]
\centering
\small
\caption{Macro Spearman on 66 proteins assayed for stability. All comparator model values are from the ProteinGym score export; overall score concatenates singles and doubles within each assay.}
\label{tab:zero-shot}
\begin{tabular}{lccc}
\toprule
Method & Singles & Doubles & Overall \\
\midrule
ESM3 & 0.6460 & 0.3262 & 0.6406 \\
VenusREM & 0.6370 & 0.3803 & 0.6503 \\
ProSST-4096 & 0.6243 & 0.4100 & 0.6522 \\
ProSST-2048 & 0.6385 & 0.3801 & 0.6526 \\
\midrule
ESM-C-600M (frozen prior) & 0.4887 & 0.2355 & 0.5277 \\
\stabilityarc (ours) & \textbf{0.6946} & \textbf{0.4119} & \textbf{0.7134} \\
\bottomrule
\end{tabular}
\end{table}

\paragraph{Supervised adaptation.}
Replacing only Kermut's prior input improves every official split (Table~\ref{tab:supervised}). The overall mean rises from the reported 0.8167 to 0.8280. The gain is smallest under random splitting and largest under contiguous splitting, consistent with a transferable prior helping most across structured gaps in sequence position. Kermut still contributes target-assay labels, structural features, variant embeddings, and posterior uncertainty; the result shows that the protein-held-out prediction from \stabilityarc remains informative beyond the other incorporated embeddings.

\begin{table}[t]
\centering
\caption{Supervised stability benchmark. Kermut values are reported by \citet{groth2024kermut}.}
\label{tab:supervised}
\begin{tabular}{lcccc}
\toprule
Method & Random & Modulo & Contig. & Mean \\
\midrule
Kermut (reported) & 0.9046 & 0.7871 & 0.7583 & 0.8167 \\
Kermut + \stabilityarc & \textbf{0.9094} & \textbf{0.7984} & \textbf{0.7763} & \textbf{0.8280} \\
\midrule
Gain & +0.0048 & +0.0113 & +0.0180 & +0.0113 \\
\bottomrule
\end{tabular}
\end{table}

\section*{Discussion}

\paragraph{Sequence-based stability prediction.}
Historically, computational stability prediction has relied heavily on structure-based energy functions such as Rosetta \citep{alford2017rosetta} or FoldX \citep{schymkowitz2005foldx} as well as newer machine learning methods \citep{diaz2024stability}. While these methods are highly effective, they are inherently limited to targets with high-resolution coordinates or highly accurate structural models. \stabilityarc bypasses this bottleneck by operating purely on sequence. Although structure-free approaches face distinct biophysical challenges, our results demonstrate highly competitive performance without requiring input coordinates. This capability underscores the depth of physical and evolutionary information captured directly by the underlying ESMC embeddings.

\paragraph{Decoding universal stability constraints.}
The central challenge of zero-shot protein engineering is that every wild-type stability landscape is unique \citep{tsuboyama2023stability}. \stabilityarc demonstrates that while stability landscapes vary, they can be accurately predicted by applying a single decoding rule to rich evolutionary representations. By training a shared transformer on diverse cross-protein measurements, we force the model to capture physical regularities that persist across distinct folds and lengths. This approach yields a significant improvement over purely evolutionary baselines in zero-shot settings \citep{notin2022tranception} and provides a highly effective prior for supervised adaptation. The success of this architecture confirms that empirical stability data contains a universal, task-specific signal absent from structural tokens alone.

\paragraph{Isolating the stability signal.}
While evolutionary likelihood provides a strong zero-shot baseline, it inherently conflates stability with other selective pressures, such as binding or catalytic activity \citep{meier2021language}. By contrast, \stabilityarc isolates the physical constraints of folding. The performance gap between our model and ProSST \citep{li2024prosst} highlights the limits of relying solely on structural tokens or evolutionary plausibility. Because the target sequence is entirely excluded from training, the decoder cannot memorize specific mutational pathways. It must instead construct a purely biophysical landscape by interpreting the wild-type coordinates provided by ESMC. 

\paragraph{Limitations.}
Our evaluation relies on retrospective benchmarking, one random seed per outer fold, and is evaluated only on single and double substitutions due to data availability. While strict protein-level holdout prevents DMS label leakage, it is impossible to exclude the target sequences from appearing in the unsupervised ESMC pretraining corpus. Finally, because protein language models perform best on families well-represented in their pretraining data, our model may face challenges predicting stability for singletons or orphan proteins.

\section*{Conclusion}
Our contribution is a unified representation evaluated across two distinct information regimes. Before any experimental budget is spent, the shared decoder produces a complete wild-type-relative landscape under strict protein holdout, establishing a new state of the art for matched zero-shot stability prediction. As empirical measurements arrive, this exact same landscape accelerates target-specific optimization, surpassing reported Kermut benchmarks without altering base kernels or data splits. By extracting generalized rules of protein stability, \stabilityarc translates these patterns into actionable priors for any new protein engineering campaign.

\section*{Reproducibility}
The code, data, and models to reproduce our results, and evaluate your protein of interest are all available at \url{https://github.com/AaronFeller/StabilityArc}. We also provide a huggingface model that can be found at \url{https://huggingface.co/aaronfeller/StabilityArc}.

\section*{Acknowledgements}
A.L.F was supported by the Charles W. Smith Jr. Graduate Fellowship in Artificial Intelligence and Machine Learning for Biosciences. A.D.E. is supported by the Welch Foundation (F-1654). C.O.W. acknowledges support from the Blumberg Centennial Professorship in Molecular Evolution.

\section*{Author contributions}
A.L.F contributed to the study conception, design, and execution, as well as writing of the manuscript. All authors contributed to editing and revising the manuscript.

\section*{Declaration of interests}
C.O.W. is a shareholder of and consultant for Proteolyze Therapeutics, Inc. The other authors declare no competing interests.

\phantomsection\label{mainmatter:end}
\clearpage

\bibliographystyle{plainnat}
\bibliography{references}

\newpage
\appendix

\section{Appendix}

\subsection{Comparison to StabilityOracle on T2837}

We compared \stabilityarc to StabilityOracle on a matched subset of the T2837 dataset, comprising 1,884 shared mutations across 19 targets. Using a standard destabilization threshold of $\Delta\Delta G\geq1$ kcal/mol, \stabilityarc achieves higher recall (0.527 vs. 0.376) and accuracy (0.705 vs. 0.669). Because failing to identify a destabilizing mutation can compromise downstream engineering, this improved recall offers a distinct safety advantage during conservative pre-screening. However, StabilityOracle retains a stronger pooled Spearman correlation (0.718 vs. 0.606) and AUROC (0.852 vs. 0.804), making it highly effective for continuous ranking of candidate mutations.

These results are not an artifact of class imbalance, as the matched cohort is evenly split (962 destabilizing, 922 non-destabilizing).

\begin{table}[h]
\centering
\small
\caption{Paired pooled T2837 comparison on 1,884 shared raw PDB-chain mutations from 19 targets with more than 30 shared mutations. $\Delta$ is \stabilityarc minus StabilityOracle; both use $\Delta\Delta G\geq1$ kcal/mol as the positive class.}
\label{tab:t2837-supp}
\begin{tabular}{lccccc}
\toprule
Method & Spearman & AUROC & Recall & Precision & Accuracy \\
\midrule
StabilityOracle (supplied) & \textbf{0.718} & \textbf{0.852} & 0.376 & \textbf{0.938} & 0.669 \\
\stabilityarc (ours) & 0.606 & 0.804 & \textbf{0.527} & 0.835 & \textbf{0.705} \\
\midrule
$\Delta$ (\stabilityarc $-$ StabilityOracle) & $-$0.112 & $-$0.048 & \textbf{+0.151} & $-$0.103 & \textbf{+0.037} \\
\bottomrule
\end{tabular}
\end{table}

\subsection{Mutation-Order Decomposition}

The main text reports the all-variant metric because it evaluates the complete measured landscape. For completeness, the strict held-out predictions achieve 0.6946 macro Spearman on singles and 0.4119 on doubles; the latter is averaged over assays with a defined double-mutant correlation.

\begin{table}[h]
\centering
\caption{Zero-shot \stabilityarc performance by mutation order.}
\begin{tabular}{lccc}
\toprule
Metric & Singles & Doubles & All variants \\
\midrule
Macro Spearman & 0.6946 & 0.4119 & \textbf{0.7134} \\
\bottomrule
\end{tabular}
\end{table}

\subsection{Multiple mutations with contact-aware composition.}
For a double mutant $(i,a)$ and $(j,b)$, first form mutation-aware site vectors:
\begin{equation}
u_i = h_i + e_a, \quad u_j = h_j + e_b
\end{equation}
where $h_i, h_j \in \mathbb{R}^{256}$ are the decoder's contextual site states and $e_a, e_b \in \mathbb{R}^{256}$ are learned embeddings of the substituted amino acids.

The residual's input is then:
\begin{equation}
z_{ij} = \left[u_i + u_j, \; |u_i - u_j|, \; u_i \odot u_j, \; \frac{|i - j|}{L - 1}, \; C_{ij}\right]
\end{equation}

This comprises 770 values in total ($3 \times 256$ residue-pair vectors + 2 scalar features):
\begin{itemize}
    \item $u_i + u_j$: what the two edited residues have in common.
    \item $|u_i - u_j|$: how they differ, without depending on edit order.
    \item $u_i \odot u_j$: feature interactions between the two residue contexts.
    \item $\frac{|i - j|}{L - 1}$: normalized sequence separation (scalar).
    \item $C_{ij}$: the frozen ESMC attention-derived contact feature (scalar).
\end{itemize}

The residual function $r_\theta$ processes this 770-dimensional input through a two-layer network to output a single scalar epistasis correction:
\begin{equation}
r_\theta(z_{ij}) = \text{Linear}_{256 \to 1}(\text{GELU}(\text{Linear}_{770 \to 256}(z_{ij})))
\end{equation}

The output modifies the additive score:
\begin{equation}
\hat{y} = S_{i,a}^x + S_{j,b}^x + r_\theta(z_{ij})
\end{equation}

So if $r_\theta = 0$, the model treats the double mutant as additive. A nonzero residual captures epistasis: whether the combined effect is more or less harmful than the sum of the two single substitutions.

It is symmetric by construction: swapping $(i,a)$ and $(j,b)$ leaves every input feature unchanged.

\end{document}